\documentclass[preprint,pra]{revtex4}%
\usepackage{amsfonts}
\usepackage{amsmath}
\usepackage{amssymb}
\usepackage{graphicx}%
\begin{document}
\title[Short title for running header]{Scattering of low- to intermediate-energy positrons from molecular hydrogen}
\author{David D. Reid, William B. Klann}
\affiliation{Department of Physics and Astronomy, Eastern Michigan University, Ypsilanti,
MI 48197}
\author{J. M. Wadehra}
\affiliation{Department of Physics and Astronomy, Wayne State University, Detroit, MI 48202}
\pacs{PACS number}

\begin{abstract}
Using a complex model potential, we have calculated the total, integrated
elastic, momentum transfer, absorption, and differential cross sections for
positrons scattered from molecular hydrogen. The widely available software
package GAUSSIAN is used to generate the radial electronic charge density of
molecule which is used to produce the interaction potentials. The quasifree
absorption potential, previously developed and used for positron-atom
scattering, is extended to positron scattering from molecular targets. It is
shown that this model potential approach produces accurate results even into
the low-energy regime.

\end{abstract}
\volumeyear{year}
\volumenumber{number}
\issuenumber{number}
\eid{identifier}
\date[Date text]{date}
\received[Received text]{date}

\revised[Revised text]{date}

\accepted[Accepted text]{date}

\published[Published text]{date}

\startpage{1}
\endpage{2}
\maketitle

\section{Introduction}

The scattering of positrons from atomic and molecular targets continues to be
an area of active investigation in both experimental and theoretical collision
studies. As the ability to produce controlled positron beams continues to be
refined, and such beams become available in more laboratories, a larger
variety of positron-gas systems are being studied experimentally with
improving results. The state of theoretical calculations in this area can be
divided into four impact-energy ($E$) regimes. These are very low energy
($E\lesssim0.1$ eV), low energy ($0.1$ eV$<E<E_{Ps}$) where $E_{Ps}$ is the
threshold for positronium formation, intermediate energy ($E_{Ps}<E<1000$ eV),
and high energy ($E>1000$ eV) regimes. Low and very low energy calculations
are typically performed at the \textit{ab initio} level rather than with model
potentials partly because in this energy regime one does not have to take into
account several inelastic channels which are complicated to handle exactly
\cite{Lino}. Furthermore, calculations using model potentials have performed
only moderately well or even poorly at lower energies because the projectile
spends more time near the target causing the results to be more sensitive to
the details of the interaction. However, the reverse is true at intermediate
energies. Because of the predominance of many inelastic processes,
particularly positronium formation, electronic excitation, and ionization,
calculations at the \textit{ab initio} level become extremely difficult. Also,
in this energy regime, high-energy approximations, such as Born-Bethe theory,
cannot yet be trusted.

We will show in this paper that use of complex model potentials can produce
accurate intermediate-energy results even for positron-molecule scattering, as
they have for the scattering of both electrons and positrons in atomic gases
\cite{RWpol}. However, despite the success of this approach for atomic targets
at intermediate energies, use of model potentials runs into difficulties that
limited their applicability to molecular targets. First, the generation of
molecular charge densities is substantially more difficult than the generation
of atomic charge densities; therefore, many of the previous calculations for
molecules employed the independent-atom model \cite{CNO, Raizada} in which the
scattering process from a molecule is treated by combining the scattering
processes from the individual atoms that make up the molecule. This approach
necessarily breaks down at lower energies, depending on the geometry of the
molecule, because when the de Broglie wavelength of the incident positrons is
on the order of the size of the bond lengths between the atoms in the molecule
they cannot possibly \textquotedblleft see\textquotedblright\ the molecule as
a set of individual atoms. Furthermore, model potentials that assume that the
electrons of the target atom can be treated as a free electron gas are not
accurate for atomic hydrogen containing only one electron. Therefore, in this
case, the independent atom approximation for molecules containing the hydrogen
atom is not expected to be very good. Second, no good model absorption
potential specifically designed for positron scattering has existed until only
recently \cite{quasi-free}. Having no viable option, previous
positron-molecule collision calculations were carried out either using model
absorption potentials that were designed for electron scattering, or modifying
those electron absorption potentials in purely empirical ways
\cite{Baluja-Jain}.

Because of the issues just described, and despite the fact that electron
scattering from molecular hydrogen is a well-studied problem, to the best of
our knowledge, there are only two published calculations of total cross
sections for positron scattering from H$_{2}$ at intermediate impact energies
\cite{Raizada, Baluja-Jain}. In this paper, we study positron-H$_{2}$
scattering in a way that addresses both of the difficulties discussed in the
previous paragraph. First, as will be discussed in more detail below, the
present calculations use \textit{molecular} charge densities to calculate the
model potentials. By doing so, we bypass all of the issues concerning use of
the independent-atom model. As a result, not only are we able to obtain good
cross section results for scattering from H$_{2}$ at intermediate impact
energies, but, surprisingly, our results are also quite good well into the
low-energy regime. Second, we demonstrate the successful extension of the
quasifree model absorption potential developed for positron-atom scattering to
the scattering of positrons from molecular targets. Using a more appropriate
positron absorption potential gives better overall results with much less need
for empiricism.

This paper is organized into four parts. Following the present introductory
remarks, we explain in section II the theoretical framework for our
calculations. First, in subsection II.A, we describe the interaction
potentials used and discuss the relevant issues concerning the extension of
the quasifree model to molecular targets. Subsection II.B is devoted to a
discussion of how we generated the molecular charge densities (and static
potential) of the target using the commercially available software GAUSSIAN
\cite{Gaussian}. The details of how these calculations were performed are then
given in subsection II.C. In section III, we present our results for total,
integrated elastic, momentum transfer, absorption, and differential cross
sections from low to intermediate impact energies. Finally, we make some
concluding remarks in section IV. Unless otherwise specified, we use atomic
units ($\hslash=e=m_{e}=1$) throughout this paper.

\section{Theory\-}

\subsection{Interaction Potentials}

In the present calculations we model the positron-target system by a complex
interaction potential, $V(r)$, that consists of three parts. These parts are
the static potential $V_{st}(r)$, the polarization potential $V_{pol}(r)$, and
the absorption potential $V_{abs}(r)$, such that%
\begin{equation}
V(r)=V_{st}(r)+V_{pol}(r)+iV_{abs}(r). \label{IP}%
\end{equation}

Each interaction potential is determined by the radially averaged electron
charge density of the target molecule, $\rho(r)$, which is obtained using the
method discussed in subsection II.B below. The static potential is given by%

\begin{equation}
V_{st}(r)=\left\langle \frac{Z}{|\mathbf{r-b|}}\right\rangle -4\pi\int
_{0}^{\infty}\frac{\rho(r^{\prime})}{r_{_{>}}}r^{\prime2}dr^{\prime},
\label{statdef}%
\end{equation}
where $Z$ is the number of protons of the target ($Z=2$ in the present case),
$\mathbf{b}$ is a vector that points from the center of the molecule to a
nucleus, and $r_{_{>}}$ is the larger of $r$ and $r^{\prime}$.

Following De Fazio \textit{et al} \cite{DeFazio}, the polarization interaction
is given, in terms of the electron density, as%

\begin{equation}
V_{pol}(r)=-D_{4}(r)\frac{\alpha_{d}}{2r^{4}}-D_{6}(r)\frac{\alpha_{q}}%
{2r^{6}}-D_{8}(r)\frac{\alpha_{o}}{2r^{8}} \label{Vpol}%
\end{equation}
where $\alpha_{d}$, $\alpha_{q}$, and $\alpha_{o}$ are the dipole, quadrupole,
and octopole polarizabilities of the target molecule, respectively. In Table
I, the values of the polarizabilities and their sources, as well as other
parameters used in these calculations are provided. In Eq. (\ref{Vpol}) the
functions $D_{2\ell+2}(r)$ are damping functions whose purpose is to guarantee
that $V_{pol}\rightarrow0$ as $r\rightarrow0$; these functions are given by%
\begin{equation}
D_{2\ell+2}(r)=\frac{\int_{0}^{r}\rho(r^{\prime})r^{\prime2\ell+2}dr^{\prime}%
}{\int_{0}^{\infty}\rho(r^{\prime})r^{\prime2\ell+2}dr^{\prime}}.
\label{damp_f}%
\end{equation}

The absorption potential used in this work is an extension of the quasi-free
model for positron-atom scattering that was given in our previous work
\cite{quasi-free}. The form of this interaction potential is%

\begin{equation}
V_{abs}=-\frac{1}{2}\rho\bar{\sigma}_{b}v, \label{Vabs}%
\end{equation}
where $v$ is the local speed of the incident positron and $\bar{\sigma}_{b}$
is the average cross section for binary collisions between the positron and
the electrons of the target molecule. One of the important aspects of the
present study is to formulate an extention of this model interaction potential
to the case of molecular targets. Besides the electron density, the only other
target-dependent quantity used in the absorption potential is the energy gap
$\Delta$. Within the quasifree binary collision model, $\Delta$ plays a dual
role as both (a) the energy gap between the initial state and the final energy
state of the originally bound electron, and (b) the lowest energy threshold
for inelasic processes. For \textit{electron}-atom scattering, these two roles
are consistent with each other if $\Delta$ is set equal to the excitation
threshold ($E_{exc}$) of the target atom. However, for \textit{positron}-atom
scattering the formation of positronium introduces another inelastic threshold
which can be lower than the threshold for excitation. As an example, for
positron scattering from alkali-metal atoms the threshold for positronium
formation ($E_{Ps}$) is zero \cite{alkalis}. In the quasifree model the
aborption cross section diverges as $\Delta\rightarrow0$. Thus, for many
positron-atom systems one has to find a reasonable choice for $\Delta$ that
will be sufficiently close to the true inelastic threshold so as to minimize
the absence of low-energy absorption in the calculations, but not so small
that cross sections begin to diverge. Our previous investigations of
positron-atom scattering \cite{quasi-free, alkalis} have suggested that the
appropriate choice for $\Delta$ is to set it equal to the lowest
\textit{nonzero} inelastic threshold.

In the case of positron scattering from molecular targets the inelasic
threshold is effectively always open because of rovibrational excitation
thresholds of the target molecules. Besides the rovibrational modes, the
possibility of the dissociation of the molecule adds an additional inelastic
process with threshold $E_{diss}$. In the derivation of the quasifree model,
the only inelastic processes that are considered are those that can result
from a binary collision between the incident positron and a target electron,
namely, electronic excitation and ionization by positron impact, and
positronium formation. Obviously, rovibrational excitation and dissociative
processes are not part of the binary collision. This would most directly
suggest that the energy gap be set equal to $E_{Ps}$. However, the above
considerations must be balanced against the other role of $\Delta$ as the
threshold at which \textit{any} inelastic scattering occurs. Therefore, in the
present study we have taken $\Delta$ to equal the average of $E_{Ps}$ and the
threshold of disscociation,
\begin{equation}
\Delta=\frac{1}{2}(E_{Ps}+E_{diss})\text{ .} \label{gap}%
\end{equation}

For positron scattering the binary collision cross section $\overline{\sigma
}_{b}$ of Eq.(\ref{Vabs}) is given by \cite{quasi-free, alkalis},
\begin{equation}
\overline{\sigma}_{b}=\frac{\pi}{\left(  \varepsilon E_{F}\right)  ^{2}%
}\left\{
\begin{array}
[c]{l}%
f\left(  0\right)  \qquad\qquad\qquad\varepsilon^{2}-\delta\leq0\\
f\left(  \sqrt{\varepsilon^{2}-\delta}\right)  \qquad0<\varepsilon^{2}%
-\delta\leq1\\
f\left(  1\right)  \qquad\qquad\qquad1<\varepsilon^{2}-\delta
\end{array}
\right.  ,
\end{equation}
where%
\begin{equation}
f(x)=\frac{2}{\delta}x^{3}+6x+3\varepsilon\ln\left(  \frac{\varepsilon
-x}{\varepsilon+x}\right)
\end{equation}
and
\begin{equation}
\delta=\frac{\Delta}{E_{F}},\qquad\varepsilon=\sqrt{\frac{E}{E_{F}}}.
\end{equation}
The quantities $E_{F}=\hslash^{2}k_{F}^{2}/2m$ and $k_{F}=(3\pi^{2}\rho
)^{1/3}$ are the Fermi energy and the Fermi wavenumber (or momentum)
corresponding to the target radial electron density $\rho$.

\subsection{The Electronic Charge Density}

In the present calculations, the electronic charge density in the hydrogen
molecule is calculated with GAUSSIAN \cite{Gaussian} using\ the full
configuration interaction method with both single and double substitutions
\cite{CI refs}. This code is now fast and readily available. Using the
\texttt{cube=density} command in GAUSSIAN, we first generated the electronic
charge density $\rho(\mathbf{r})$ on a sufficiently large three-dimensional
cubic grid to cover the needed range of the calculation with a step size of
0.04 a$_{0}$ in each direction. By interpolation \cite{numrec}, we then
obtained values of $\rho(\mathbf{r})$ over the surface of a sphere of radius
$r$ centered upon the geometric center of the molecule; Fig. 1 illustrates
this procedure. For visual clarity, Fig. 1 only shows points on a plane; in
fact, the symmetry of H$_{2}$ only requires generation of $\rho(\mathbf{r})$
over one quadrant of such a plane. The value of the radial charge density at
$r$ is then calculated by numerical integration%
\begin{equation}
\rho(r)=\frac{1}{4\pi}\int_{0}^{2\pi}\int_{0}^{\pi}\rho(\mathbf{r})\sin
\theta\text{\/\thinspace}d\phi d\theta. \label{radial cd}%
\end{equation}
In this manner, values of $\rho(r)$ are calculated for every value of $r$
needed in the integration of the radial Schr\"{o}dinger equation to be
discussed in the next subsection.

\subsection{Calculations}

For the spherically symmetric potential of Eq. (\ref{IP}) the scattering
process is symmetric about the direction of the incident positron. The
solution $u_{\ell}(r)$, therefore, is generated by the radial Schr\"{o}dinger
equation (in atomic units)
\begin{equation}
\left[  \frac{d^{2}}{dr^{2}}-\frac{\ell(\ell+1)}{r^{2}}+2\left[
E-V(r)\right]  \right]  u_{\ell}(r)=0 \label{Schrodinger}%
\end{equation}
where $E=\hslash^{2}k^{2}/2m$ is the impact energy of the collision and $\ell$
is the angular momentum quantum number which also represents the order of the
partial wave \cite{Schiff}.

Equation (\ref{Schrodinger}) is integrated out to a distance of 10 bohr radii
from the center of the molecule via the Numerov technique \cite{numerov}. The
first 51 ($\ell_{\max}=50$) phase shifts are calculated exactly by comparing
$u_{_{\ell}}$, the radial wave function of the target plus positron system, at
two adjacent points $r$ and $r_{_{+}}=r+h$:%

\begin{equation}
\tan\left(  \delta_{\ell}\right)  =\frac{r_{_{+}}u_{_{\ell}}(r)j_{_{\ell}%
}(kr_{_{+}})-ru_{_{\ell}}(r_{_{+}})j_{_{\ell}}(kr)}{ru_{_{\ell}}(r_{_{+}%
})n_{_{\ell}}(kr)-r_{_{+}}u_{_{\ell}}(r)n_{_{\ell}}(kr_{_{+}})},
\label{phases}%
\end{equation}
where $h$ is the step size ($h=$ $0.00075$ a$_{0}$) of the calculation, and
$j_{_{\ell}}$ and\ $n_{_{\ell}}$ are the spherical Bessel and Neumann
functions evaluated using the algorithm of Gillman and Fiebig \cite{Gillman}.

The scattering amplitude is obtained from the phase shifts by%

\begin{equation}
f(\theta)=\frac{1}{2ik}\sum_{\ell=0}^{\ell_{\max}}(2\ell+1)(\exp
(2i\delta_{_{\ell}})-1)P_{\ell}(cos\theta)+f_{4}(\theta)+f_{6}(\theta
)+f_{8}(\theta). \label{scatt amp}%
\end{equation}
The functions $f_{4}$, $f_{6}$, and $f_{8}$ are the higher-$\ell$
contributions from the Born phase shifts for the dipole ($\thicksim1/r^{4}$),
quadrupole ($\thicksim1/r^{6}$), and octopole ($\thicksim1/r^{8}$)\ parts of
the asymptotic polarization potential, respectively. The closed form
expressions for these functions are \cite{Wadehra-Nahar}%

\begin{equation}
f_{4}(\theta)=-\pi k\alpha_{d}\left(  \frac{\sin(\theta/2)}{2}+\sum_{\ell
=0}^{\ell_{\max}}\frac{P_{\ell}(\cos\theta)}{(2\ell+3)(2\ell-1)}\right)  ,
\end{equation}%
\begin{equation}
f_{6}(\theta)=-3\pi k^{3}\alpha_{q}\left(  -\frac{\sin^{3}(\theta/2)}{18}%
+\sum_{\ell=0}^{\ell_{\max}}\frac{P_{\ell}(\cos\theta)}{(2\ell+5)(2\ell
+3)(2\ell-1)(2\ell-3)}\right)  ,
\end{equation}
and%
\begin{equation}
f_{8}(\theta)=-10\pi k^{5}\alpha_{o}\left(  \frac{\sin^{5}(\theta/2)}%
{450}+\sum_{\ell=0}^{\ell_{\max}}\frac{P_{\ell}(\cos\theta)}{(2\ell
+7)(2\ell+5)(2\ell+3)(2\ell-1)(2\ell-3)(2\ell-5)}\right)  .
\end{equation}

Once the scattering amplitude is known, the various cross sections can be
determined. The total cross sections which include both elastic and inelastic
scattering, are obtained from the forward scattering amplitude by%
\begin{equation}
\sigma_{tot}=\frac{4\pi}{k}\operatorname{Im}\left[  f\left(  0\right)
\right]  \text{ .} \label{TCS}%
\end{equation}
The cross sections for elastic scattering are found by integrating the
scattering amplitude%
\begin{equation}
\sigma_{elas}=2\pi\int_{0}^{\pi}\left\vert f(\theta)\right\vert ^{2}\sin
\theta\,d\theta\text{ .} \label{ICS}%
\end{equation}
The absorption cross sections (the cross section for inelastic scattering) are
determined by the difference%
\begin{equation}
\sigma_{abs}=\sigma_{tot}-\sigma_{elas}\text{ .} \label{ACS}%
\end{equation}
The differential cross sections for the angular distribution of the scattered
wave are given by%

\begin{equation}
\frac{d\sigma}{d\Omega}=\left\vert f(\theta)\right\vert ^{2}\text{ .}
\label{DCS}%
\end{equation}
Finally, the momentum transfer cross sections are found using%
\begin{equation}
\sigma_{mom}=2\pi\int_{0}^{\pi}\left(  1-\cos\theta\right)  \left\vert
f(\theta)\right\vert ^{2}\sin\theta\,d\theta\text{ .} \label{MTCS}%
\end{equation}

\section{Results}

Figure 2 shows the present results of the total cross sections for the
scattering of positrons by H$_{2}$ compared with several experimental
measurements. To the best of our knowledge, no other theoretical calculations
of total cross sections have been able to predict the stucture in this curve
over as large a range of positron energies as in the present calculations.
These structures extending across the low- to intermediate-energy ranges are
accurately reproduced. The present results corrrectly predict the local
minimum in the low-energy regime near 4 eV and the local maximum in the
intermediate-energy regime near 25 eV. In the range of around $9$ eV to $11$
eV the present results stray outside of the error bars, overestimating the
experimental values. However, in this connection, it should be noted that
cross section measurements are expected to be underestimated due to the
inability to discriminate projectiles elastically scattered through small
angles \cite{Hoffman}. To get the best indication of the quality of the
present calculations, the error bars shown in Fig. 2 are the "maximum errors"
as reported in Refs. \cite{Hoffman, Zhou} and not just the statistical
uncertainties. Error bars for the other experimental data are not shown as the
errors reported were not of comparable detail.

In Fig. 3, we show our absorption cross section results compared to estimates
based on various measurements. The experimental points are a combination of
different experiments for measurements made at common, or nearly common,
impact energies. The present results show good ageement with the experimental
cross sections in the region of overlap. The fact that our results
overestimate the experimental points at every energy is to be expected because
the ionization cross sections are only for first ionizations, the excitation
cross sections only account for excitations to the B$^{1}\Sigma$ state, and
there is no experimental data added for other processes (although they are
expected to be small at these energies). As one would expect, the absorption
cross sections are quite sensitive to the absorption potential; the fact that
we have such good results for this partial cross section, confirms the
applicability of the quasifree model for molecular targets.

Our demonstration, in Figs. 1 and 2, that the present total and absorption
cross sections are good also confirms the quality of our integrated elastic
cross sections at intermediate energies. In Table II, we provide the values of
our differential, integrated elastic, and momentum transfer cross sections at
intermediate impact energies. As mentioned in sec. I, we can also claim that
the present model potential results are reliable well into the low-energy
regime. This is confirmed partly by the quality of the low-energy total cross
sections in Fig. 2. However, a much more stringent test is made by
differential cross sections. To date, there are no measurements of
differential cross sections for positron scattering from H$_{2}$. Thus, in
Fig. 4, we compare our present low-energy differential cross sections against
the \textit{ab initio} calculations of Lino \textit{et al} using the Schwinger
multichannel method \cite{Lino}. Despite the fact that, at small scattering
angles, our calculations show a slight dip, the present results show excellent
agreement with their calculations at every energy for which a comparison has
been made.

\section{Conclusions}

In the present calculations, we have presented calculations of scattering
cross sections for positrons scattered from H$_{2}$. Using a single model
potential approach, we have presented accurate total cross sections through
both the low- and intermediate-energy regimes correctly matching the detailed
structure in this curve. To the best of our knowledge, this is the first
theoretical calculation to achieve this feat. We have also demonstrated that,
with a very minor modification, the positron quasifree absorption potential
can perform equally well, or better, for scattering in molecular gases as it
has in atomic gases. Furthermore, we have introduced a simple scheme for
obtaining accurate molecular charged densities using GAUSSIAN that can be
applied to almost any molecule bypassing the need for the independent atom model.

\begin{acknowledgments}
We wish to thank H. B. Schlegel and M. C. Milletti for recommending the use of
GAUSSIAN for calculating molecular charge densities. We also acknowledge G.
Maroulis and D. M. Bishop for advise concerning the polarizabilities of
H$_{2}$. The assistance of C. M. Surko and J. P. Marler with their values of
the excitation cross sections is greatly appreciated. Completion of this
research was made possible by a Spring-Summer research award from Eastern
Michigan University.
\end{acknowledgments}

\newpage

\bigskip Table I. The values of various parameters used in this work and their sources.

\begin{tabular}
[c]{p{1.5in}p{1.5in}p{1.5in}}\hline
Quantity & Value & Source\\\hline\hline
bond length & 1.401 a$_{0}$ & \cite{Interatomic}\\
$\alpha_{d}$ & 5.18 a$_{0}^{3}$ & \cite{Bishop}\\
$\alpha_{d}$ & 7.88 a$_{0}^{5}$ & \cite{Bishop}\\
$\alpha_{o}$ & 3.85 a$_{0}^{7}$ & \cite{Maroulis}\\
$E_{Ps}$ & 8.63 eV & This work\\
$E_{diss}$ & 4.52 eV & \cite{Huber}\\
$\Delta$ & 6.57 eV & This work\\\hline
\end{tabular}

\newpage\newpage Table II. Differential, integrated elastic, and momentum
transfer cross sections at selected intermediate impact energies (in atomic
units). The notation $a$ ($b$) means $a\times10^{b}$.%

\begin{tabular}
[c]{cp{0.83in}p{0.83in}p{0.83in}p{0.83in}p{0.83in}p{0.83in}}\hline\hline
Angle (deg.) & 50 eV & 100 eV & 200 eV & 300 eV & 400 eV & 500 eV\\\hline
0 & 5.88 ($0$) & 5.23 ($0$) & 3.51 ($0$) & 2.92 ($0$) & 2.38 ($0$) & 2.26
($0$)\\
10 & 4.40 ($0$) & 3.44 ($0$) & 2.00 ($0$) & 1.69 ($0$) & 1.32 ($0$) & 1.06
($0$)\\
20 & 2.62 ($0$) & 1.69 ($0$) & 7.63 ($-1$) & 6.01 ($-1$) & 3.92 ($-1$) & 2.88
($-1$)\\
30 & 1.38 ($0$) & 6.67 ($-1$) & 1.99 ($-1$) & 1.57 ($-1$) & 9.85 ($-2$) & 6.51
($-2$)\\
40 & 6.09 ($-1$) & 2.04 ($-1$) & 3.79 ($-2$) & 5.22 ($-2$) & 3.72 ($-2$) &
2.96 ($-2$)\\
50 & 2.20 ($-1$) & 5.04 ($-2$) & 5.68 ($-3$) & 2.98 ($-2$) & 2.57 ($-2$) &
2.28 ($-2$)\\
60 & 6.43 ($-2$) & 1.05 ($-2$) & 8.76 ($-4$) & 2.41 ($-2$) & 2.27 ($-2$) &
2.13 ($-2$)\\
70 & 1.44 ($-2$) & 2.42 ($-3$) & 4.03 ($-4$) & 2.19 ($-2$) & 2.16 ($-2$) &
2.10 ($-2$)\\
80 & 2.68 ($-3$) & 1.14 ($-3$) & 3.41 ($-4$) & 2.08 ($-2$) & 2.13 ($-2$) &
2.13 ($-2$)\\
90 & 1.22 ($-3$) & 8.81 ($-4$) & 2.67 ($-4$) & 2.03 ($-2$) & 2.14 ($-2$) &
2.18 ($-2$)\\
100 & 1.40 ($-3$) & 6.88 ($-4$) & 2.02 ($-4$) & 2.01 ($-2$) & 2.17 ($-2$) &
2.23 ($-2$)\\
110 & 1.34 ($-3$) & 5.00 ($-4$) & 1.52 ($-4$) & 2.01 ($-2$) & 2.20 ($-2$) &
2.26 ($-2$)\\
120 & 1.10 ($-3$) & 3.65 ($-4$) & 1.12 ($-4$) & 2.02 ($-2$) & 2.23 ($-2$) &
2.29 ($-2$)\\
130 & 8.23 ($-4$) & 2.69 ($-4$) & 8.55 ($-5$) & 2.03 ($-2$) & 2.26 ($-2$) &
2.30 ($-2$)\\
140 & 5.67 ($-4$) & 2.07 ($-4$) & 6.92 ($-5$) & 2.04 ($-2$) & 2.27 ($-2$) &
2.30 ($-2$)\\
150 & 3.92 ($-4$) & 1.75 ($-4$) & 5.71 ($-5$) & 2.06 ($-2$) & 2.29 ($-2$) &
2.31 ($-2$)\\
160 & 2.93 ($-4$) & 1.60 ($-4$) & 4.85 ($-5$) & 2.07 ($-2$) & 2.29 ($-2$) &
2.30 ($-2$)\\
170 & 2.44 ($-4$) & 1.48 ($-4$) & 4.42 ($-5$) & 2.07 ($-2$) & 2.30 ($-2$) &
2.30 ($-2$)\\
180 & 2.17 ($-4$) & 1.35 ($-4$) & 4.06 ($-5$) & 2.07 ($-2$) & 2.30 ($-2$) &
2.31 ($-2$)\\
$\sigma_{elas}$ & 3.37 ($0$) & 1.94 ($0$) & 8.80 ($-1$) & 9.59 ($-1$) & 7.64
($-1$) & 6.41 ($-1$)\\
$\sigma_{mom}$ & 3.92 ($-1$) & 1.57 ($-1$) & 4.79 ($-2$) & 2.95 ($-1$) & 3.01
($-1$) & 2.96 ($-1$)\\\hline\hline
\end{tabular}

\newpage

\section{\bigskip Figure Captions}

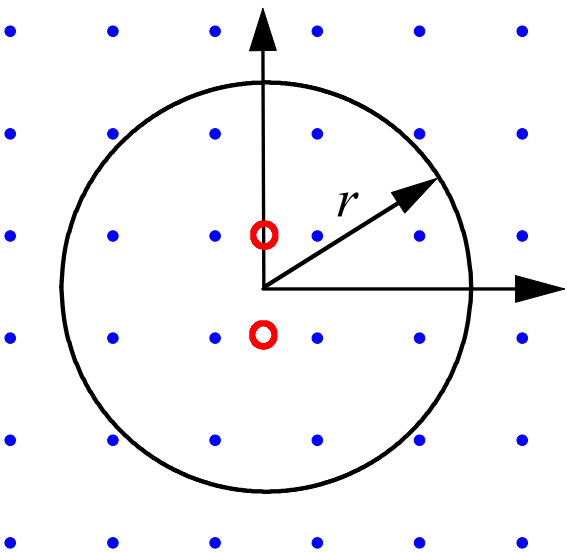
Figure 1. Points on a plane of the configuration used to generate the radial
electron charge density of H$_{2}$. The two small circles present the protons
in the hydrogen molecule. The large circle represents points on the surface of
a sphere of radius $r$. The dots represent points at which $\rho(\mathbf{r})$
is determined by GAUSSIAN and these values are used to calculate
$\rho(\mathbf{r})$ at 40,000 points on the sphere by interpolation. The radial
charge density is then determined using Eq. (\ref{radial cd}).

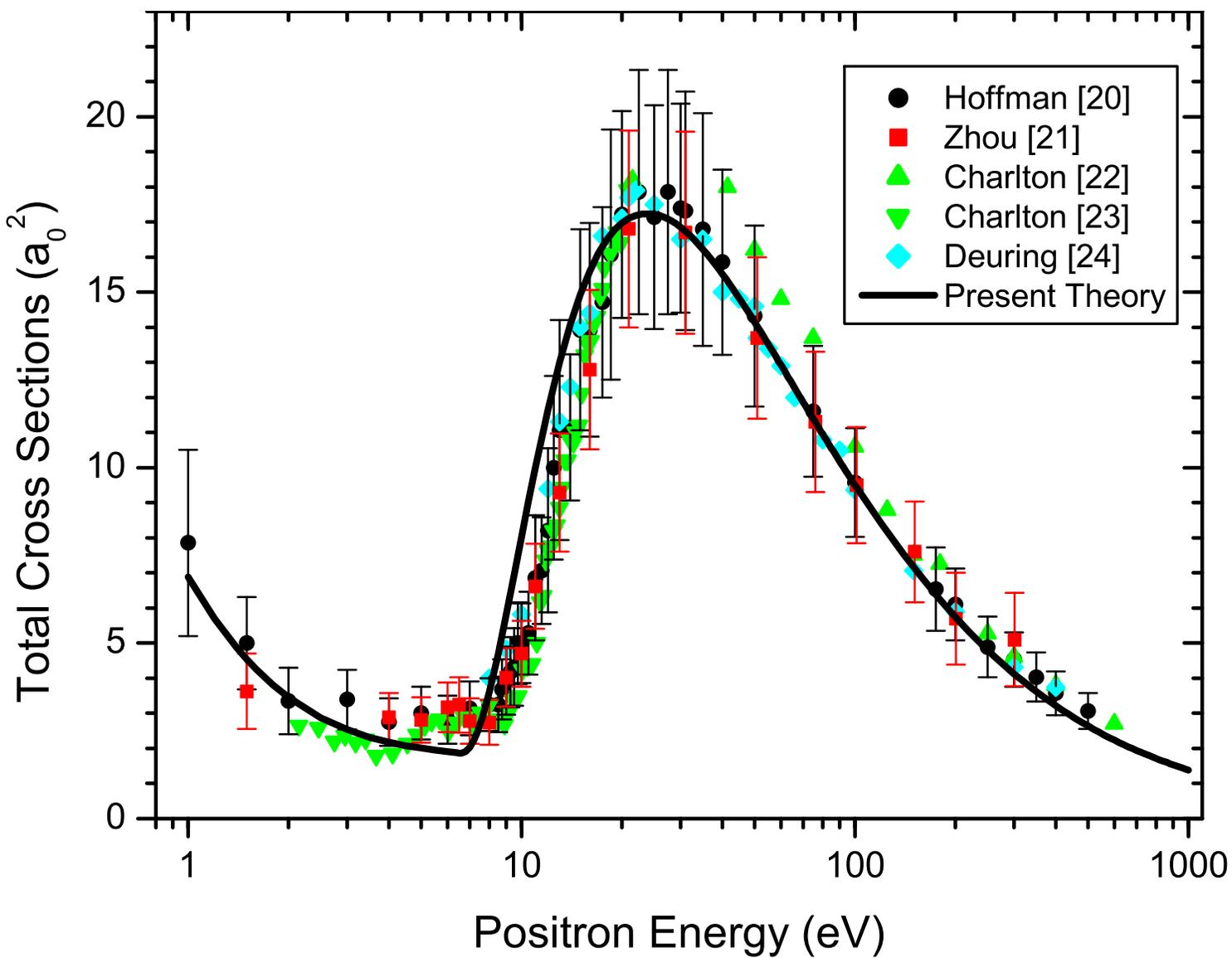
Figure 2. The present total cross sections for the scattering of low to
intermediate energy positrons by H$_{2}$ compared with several experimental
measurements. The error bars are the \textquotedblleft maximum
error\textquotedblright\ as reported by the relevant authors.

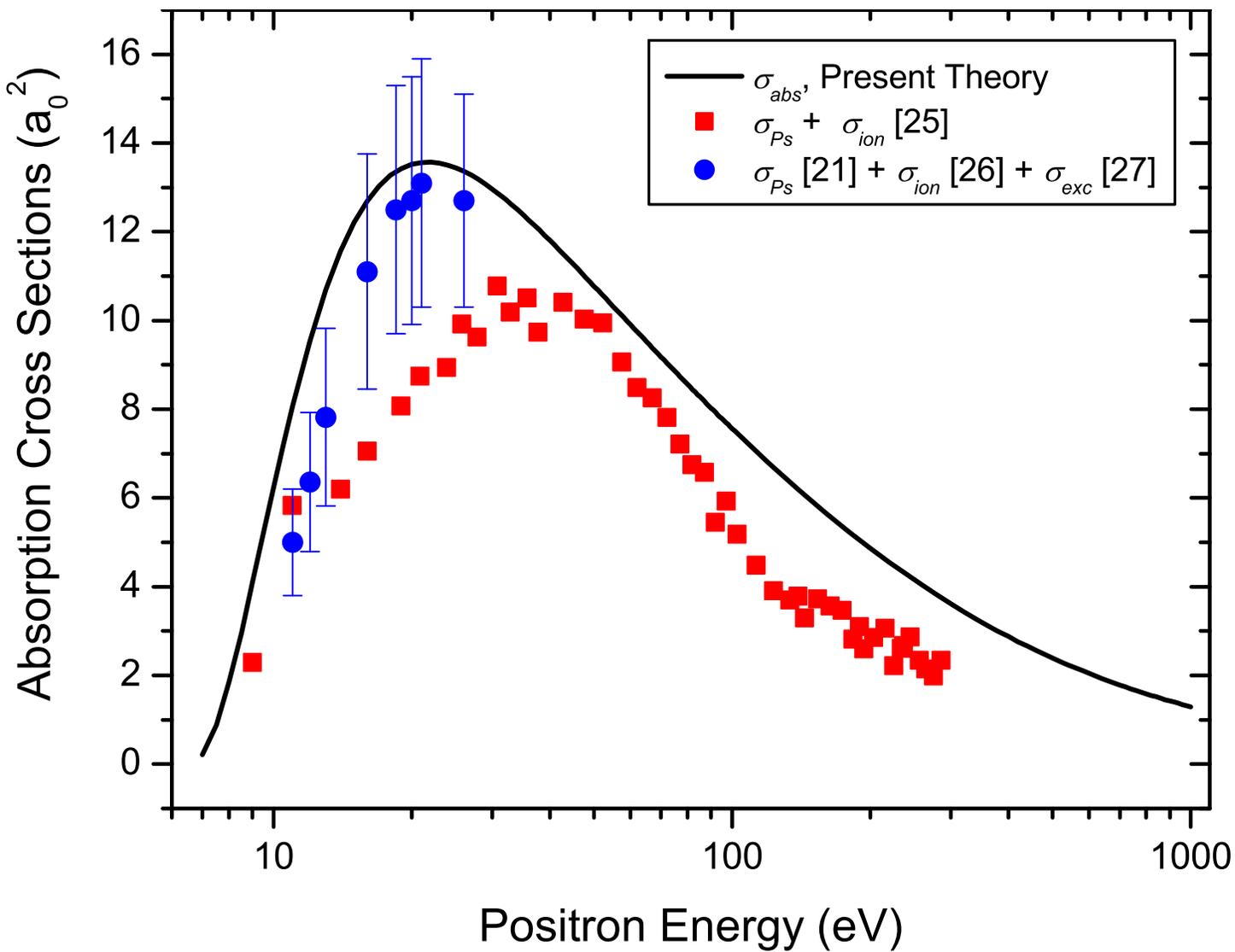
Figure 3. The present absorption cross sections for the scattering of
positrons by H$_{2}$ compared with experimental results. The experimental
results are a summation of partial cross section measurements from different
experiments. These partial cross sections are for positronium formation by
Zhou \textit{et al} \cite{Zhou}, first ionization by Maxom \textit{et al}
\cite{Maxom}, and excitation to the $B^{1}\Sigma$ state by Sullivan \textit{et
al} \cite{Sullivan}.

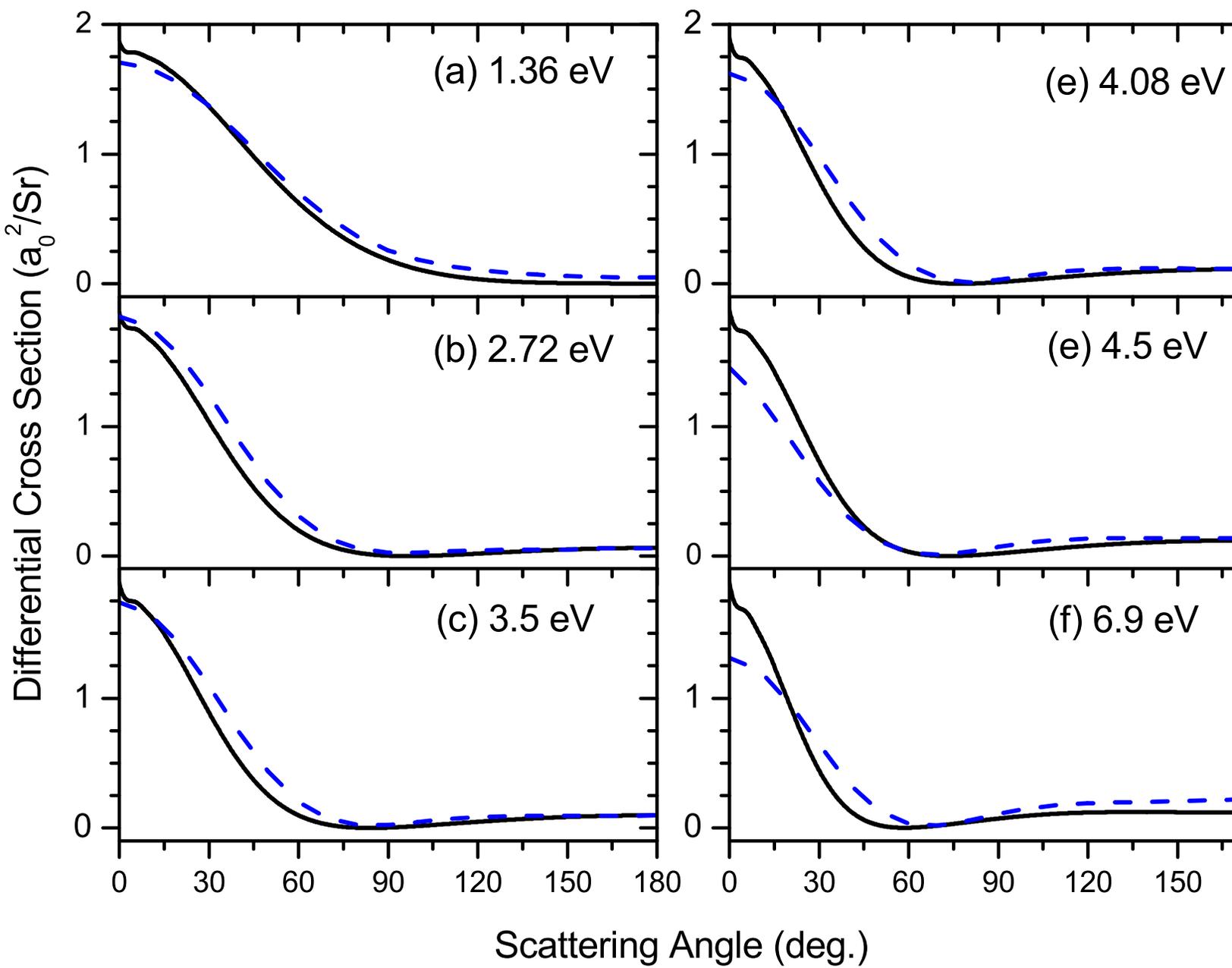
Figure 4. The present low-energy differential cross sections for the
scattering of positrons by H$_{2}$ compared with the \textit{ab initio}
calculations of Lino \textit{et al} \cite{Lino}. The positron energy ranges
from 1.36 eV to 6.9 eV.


\begin{thebibliography}{99}                                                                                               %
\newpage

\bibitem {Lino}J. L. S. Lino, J. S. E. Germano, E. P. da Silva, and M. A. P.
Lima, Phys. Rev. A \textbf{58}, 3502 (1998).

\bibitem {RWpol}D. D. Reid and J. M. Wadehra, Phys. Rev. A \textbf{50}, 4859 (1994).

\bibitem {CNO}D. D. Reid and J. M. Wadehra, Chem. Phys. Lett. \textbf{311},
385 (1999).

\bibitem {Raizada}R. Raizada and K. L. Baluja, Phys. Rev. A \textbf{55}, 1533 (1997).

\bibitem {quasi-free}D. D. Reid and J. M. Wadehra, J. Phys. B \textbf{29},
L127 (1996); B \textbf{30}, 2318 (1997).

\bibitem {Baluja-Jain}K. L. Baluja and A. Jain, Phys. Rev. A \textbf{45}, 7838 (1992).

\bibitem {Gaussian}Gaussian 98 (Revision A.11), M. J. Frisch, G. W. Trucks, H.
B. Schlegel, G. E. Scuseria, M. A. Robb, J. R. Cheeseman, V. G. Zakrzewski, J.
A. Montgomery, R. E. Stratmann, J. C. Burant, S. Dapprich, J. M. Millam, A. D.
Daniels, K. N. Kudin, M. C. Strain, O. Farkas, J. Tomasi, V. Barone, M. Cossi,
R. Cammi, B. Mennucci, C. Pomelli, C. Adamo, S. Clifford, J. Ochterski, G. A.
Petersson, P. Y. Ayala, Q. Cui, K. Morokuma, D. K. Malick, A. D. Rabuck, K.
Raghavachari, J. B. Foresman, J. Cioslowski, J. V. Ortiz, B. B. Stefanov, G.
Liu, A. Liashenko, P. Piskorz, I. Komaromi, R. Gomperts, R. L. Martin, D. J.
Fox, T. Keith, M. A. Al-Laham, C. Y. Peng, A. Nanayakkara, C. Gonzalez, M.
Challacombe, P. M. W. Gill, B. G. Johnson, W. Chen, M. W. Wong, J. L. Andres,
M. Head-Gordon, E. S. Replogle and J. A. Pople, Gaussian, Inc., Pittsburgh PA, 1998.

\bibitem {DeFazio}D. De Fazio, F. A. Gianturco, J. A. Rodriguez-Ruiz, and K.
T. Tang, J. Phys. B \textbf{27}, 303 (1994).

\bibitem {Interatomic}H. J. M. Bowen, J. Donohue, D. G. Jenkin, O. Kennard, J.
Wheatley, and D. H. Whiffen, \textquotedblleft\textit{Tables of Interatomic
Distances and Configuration in Molecules and Ions},\textquotedblright\ The
Chemical Society (London, 1958).

\bibitem {Bishop}D. M. Bishop, J. Pipin, and S. M. Cybulski, Phys. Rev. A.
\textbf{43}, 4845 (1991).

\bibitem {Maroulis}G. Maroulis and D. M. Bishop, Chem. Phys. Lett.
\textbf{128}, 462 (1986).

\bibitem {Huber}K. P. Huber and G. Herzberg, \textquotedblleft%
\textit{Molecular Spectra and Molecular Structure Constants of Diatomic
Molecules},\textquotedblright\ Van Nostrand Reinhold (New York, 1979).

\bibitem {alkalis}D. D. Reid and J. M. Wadehra, Phys. Rev. A \textbf{57}, 2583 (1998).

\bibitem {CI refs}J. A. Pople, R. Seeger, and R. Krishnan, Int. J. Quant.
Chem. Symp. \textbf{11}, 149 (1977); R. Krishnan, H. B. Schlegel, and J. A.
Pople, J. Chem. Phys. \textbf{72}, 4654 (1980); K. Raghavachari and J. A.
Pople, Int. J. Quant. Chem. \textbf{20}, 167 (1981).

\bibitem {numrec}W. H. Press, S. A. Teukolsky, W. T. Vetterling, and B. P.
Flannery, \textquotedblleft\textit{Numerical Recipes in Fortran: The Art of
Scientific Computing},\textquotedblright\ 2nd. ed., Cambridge University Press
(Cambridge, 1992).

\bibitem {Schiff}L. I. Schiff, \textquotedblleft\textit{Quantum Mechanics}%
,\textquotedblright\ 3rd. ed., McGraw-Hill (New York, 1968), Ch. 5.

\bibitem {numerov}K. Smith, \textquotedblleft\textit{The Calculation of Atomic
Collision Processes},\textquotedblright\ John Wiley \& Sons (New York, 1971).

\bibitem {Gillman}E. Gillman and H. R. Fiebig, Comput. Phys. \textbf{2}, 62 (1988).

\bibitem {Wadehra-Nahar}J. M. Wadehra and S. N. Nahar, Phys. Rev. A
\textbf{36}, 1458 (1987).

\bibitem {Hoffman}K. R. Hoffman, M. S. Dababneh, Y. -F. Hsieh, W. E. Kauppila,
V. Pol, J. H. Smart, and T. S. Stein, Phys. Rev. A \textbf{25}, 1393 (1982).

\bibitem {Zhou}S. Zhou, H. Li, W. E. Kauppila, C. K. Kwan, and T. S. Stein,
Phys. Rev. A \textbf{55}, 361 (1997).

\bibitem {Charlton80}M. Charlton, T. C. Griffith, G. R. Heyland, and G. L.
Wright, J. Phys. B \textbf{13}, L353 (1980).

\bibitem {Charlton83}M. Charlton, T. C. Griffith, G. R. Heyland, and G. L.
Wright, J. Phys. B \textbf{16}, 323 (1983).

\bibitem {Deuring}A. Deuring, K. Floeder, D. Fromme, W. Raith, A. Schwab, G.
Sinapius, P. W. Zitzewitz, and J. Krug, J. Phys. B \textbf{16}, 1633 (1983).

\bibitem {Fromme}D. Fromme, G. Kruse, W. Raith, and G. Sinapius, J. Phys. B
\textbf{21}, L261 (1988).

\bibitem {Ashley}P. Ashley, J. Maxom, and G. Laricchia, Phys. Rev. Lett.
\textbf{77}, 1250 (1996).

\bibitem {Sullivan}J. P. Sullivan, J. P. Marler, S. J. Gilbert, S. J. Buckman,
and C. M. Surko, Phys. Rev. Lett. \textbf{87}, 073201 (2001).
\end{thebibliography}
\end{document}